\documentclass[twocolumn,showpacs,preprintnumbers,amsmath,amssymb]{revtex4}

\usepackage{graphicx}
\usepackage{dcolumn}
\usepackage{bm}
\usepackage{dsfont}
\begin{document}  

\title{Dual quark condensate and dressed Polyakov loops}

\author{Erek Bilgici$^a$, Falk Bruckmann$^b$, 
Christof Gattringer$^a$ and Christian Hagen$^b$
\vspace{2mm}}

\affiliation{$^a$Institut f\"ur Physik, FB Theoretische Physik, 
Universit\"at Graz, A-8010 Graz, Austria \\
$^b$Institut f\"ur Theoretische Physik, Universit\"at Regensburg,
D-93040 Regensburg, Germany}

\begin{abstract}
We construct a new order parameter for finite temperature QCD by considering the
quark condensate for U(1)-valued temporal boundary conditions for the fermions.
Fourier transformation with respect to the boundary condition defines the dual
condensate. This quantity corresponds to an equivalence class of Polyakov loops,
thereby being an order parameter for the center symmetry.  We explore the duality
relation between the quark
condensate and these dressed Polyakov loops numerically, using quenched lattice QCD
configurations below and above the QCD phase transition. It is demonstrated that the
Dirac spectrum responds differently to changing the boundary condition, in a manner
that reproduces the expected Polyakov loop pattern. We find the dressed Polyakov
loops to be dominated by the lowest Dirac modes, in contrast to thin Polyakov loops
investigated earlier. 
\end{abstract}
\maketitle

\vskip4mm
\noindent
{\bf Introductory remarks}
\vskip1mm
\noindent
Understanding the nature of confinement has been a challenging task for many years.
Several scenarios with different candidates for the  relevant gluonic excitations were
proposed, but no closed picture has emerged yet   (it is even still debated whether
confinement is predominantly an infrared or an ultraviolet phenomenon). Also a 
connection of confinement to chiral symmetry and its breaking has been conjectured, but
not been shown either.   

In recent work \cite{paper1,paper2,paper3} we have explored the idea of connecting 
quantities sensitive to confinement to spectral sums for Dirac and covariant Laplace 
operators. These ideas were developed further in \cite{wipf,wipf2,paper4} where it was shown that 
also other quantities such as quark propagators and heat kernels may be turned into
order parameters  for the breaking of center symmetry. Spectral sums provide a
natural decomposition  into infrared (IR) and ultraviolet (UV) parts and allow one to
analyze their respective role in  confinement, as studied numerically using quenched
\cite{paper2,paper3,wipf,wipf2,paper4} and dynamical \cite{soeldner} lattice configurations.

In this letter we build on those results and develop a new order parameter for center
symmetry. In particular we Fourier transform the quark condensate (that turns into the
chiral condensate in the massless limit) with respect to a U(1)-valued temporal
boundary condition for the fermions, which we parameterize with a phase 
$\varphi \in [0,2\pi)$. We show that this duality transformation turns the
quark condensate into the expectation value of an equivalence class of Polyakov loops
which all have the same winding number $n \in \mathds{Z}$. The winding number $n$ is 
the conjugate variable to the phase $\varphi$. To the equivalence class of loops with 
winding number $n = 1$, which transforms under center transformations in the same way as 
the conventional thin Polyakov loop, we refer to as the "dressed Polyakov loop".

Since for pure gauge theory the deconfinement transition can be understood as
spontaneous breaking of the center symmetry \cite{znbreaking}, the dressed Polyakov
loop is an order parameter for confinement in pure gauge theory. The center
transformation property  of the dressed Polyakov loop is independent of the quark
mass parameter $m$. It is particularly interesting that the mass parameter $m$ may
be used to relate the chiral  condensate and the conventional thin Polyakov loop.
In the limit where the quark mass parameter $m$ is sent to infinity, the dressed 
Polyakov loop reduces to the thin loop that winds once. Conversely, for the more 
interesting limit $m \rightarrow 0$ we recover the dual of the conventional chiral 
condensate. We stress that this dual has a vanishing value below $T_c$, while it 
acquires a non-zero value above the transition.  

The dressed Polyakov loop can be written as a spectral sum for Dirac eigenvalues. Using
quenched SU(3) gauge configurations below and above the critical temperature, we  show
that dressed loops are IR dominated  (in contrast to conventional thin Polyakov loops
\cite{paper2}) and that the deconfinement transition is signaled by a changing
dependence of the IR modes on the boundary condition.

\vskip4mm
\noindent
{\bf Dual quark condensate}
\vskip1mm
\noindent
We begin our discussion with recalling the definition of the
chiral  condensate $\Sigma$. The starting point is the scalar expectation value 
$\Sigma(m,V)$ of the fermion bilinear $\overline{\psi} \psi$
evaluated at finite volume $V$ and mass $m$,
\begin{equation}
\Sigma(m,V)  =   
- \! \int \! \frac{d^4x}{V} \Big\langle \overline{\psi}(x) \psi(x)\! \Big\rangle 
 = \frac{1}{V}
 \Big\langle \! \mbox{Tr} \!\left[ ( m + D )^{-1} \right]\!\! \Big\rangle_G .
\label{condensate}
\end{equation}
In the second step the fermions were integrated out and the remaining expectation value
$\langle ... \rangle_G$ is the path integral over the gauge fields with gauge action
and fermion determinant included in the weight factor. $D$ denotes the Dirac operator
at vanishing quark mass. We refer to (\ref{condensate}) as the "quark condensate". The
chiral condensate, i.e., the proper order parameter for chiral symmetry breaking is
obtained through a double limit, where first the 4-volume $V$ is sent to infinity and
then the quark mass $m$ to zero: 
$\Sigma = \lim_{m\rightarrow 0} \, \lim_{V \rightarrow
\infty} \, \Sigma(m,V).$

An important result, that we will return to later, is the Banks-Casher formula
\cite{baca} which relates the chiral condensate to the density $\rho(0)$ of eigenvalues
at  the origin: $\Sigma = \pi  \rho(0)$. Below the critical temperature $T_c$, where
chiral symmetry is broken ($\Sigma \neq 0$), the eigenvalue density $\rho(0)$ at the
origin is  non-vanishing, while above $T_c$ the Dirac spectrum develops a gap,
$\rho(0)$ vanishes and chiral symmetry is restored ($\Sigma = 0$).  Chiral symmetry
breaking is a feature of full QCD, i.e., massless fermions are taken into account.
However, the above discussed mechanism of a non-vanishing  spectral density $\rho(0)$
below $T_c$ and a spectral gap above $T_c$ is known to hold also for the quenched case,
i.e., pure gauge theory (see, e.g., \cite{bacanumerics} for a lattice study of this
property). 

We now define the dual quark condensate which we will later identify as
an order parameter for center symmetry.   We work in a finite Euclidean volume with 
temporal  extent $\beta = 1/k_BT$. For the fermion fields $\psi$ we use the generalized
temporal boundary conditions $\psi(\vec{x},\beta) = e^{i\varphi} \,
\psi(\vec{x},0)$. The canonical choice is
anti-periodic, i.e., $\varphi = \pi$, while here we allow for arbitrary  values
$\varphi \in [0,2\pi]$.   For the spatial directions the fermions  obey periodic
boundary conditions, and the gauge fields  are periodic in all four directions. 

Now the quark condensate is considered for an arbitrary  boundary angle $\varphi$
indicated by a subscript for the Dirac operator. The "dual quark condensate"
$\widetilde{\Sigma}_n$ is defined  as the Fourier transform  with respect to
$\varphi$,  
\begin{equation}
\widetilde{\Sigma}_n (m,V) \, = \, 
\int_0^{2\pi} \! \frac{d\varphi}{2\pi} \,
 \frac{e^{-i\varphi n}}{V} \,
 \Big\langle \mbox{Tr} \left[ ( m + D_\varphi )^{-1} \right]
 \Big\rangle_G ,
\label{dualcondensate}
\end{equation}
where the index $n$ is  an integer. This gauge invariant quantity, 
in particular the case $n=1$, will be investigated analytically and numerically now.

\vskip4mm
\noindent
{\bf Relation to dressed Polyakov loops}
\vskip1mm
\noindent
Using the lattice regularization, we now discuss the relation of the dual condensate to
equivalence classes of Polyakov loops (dressed Polyakov loops). To be specific, we use
the staggered lattice Dirac operator (other lattice Dirac operators give equivalent
results) 
\begin{equation}
D_{xy} = 
\sum_{\mu = 1}^4 \! \frac{\eta_\mu(x)}{2a} \Big[ U_\mu(x) \delta_{x+\hat{\mu},y} - 
U_\mu(x\!-\!\hat{\mu})^\dagger \delta_{x-\hat{\mu},y} \Big],
\label{staggeredD}
\end{equation}
with the staggered sign function   {$\eta_\mu(x)\!= \!(-1)^{x_1+  \, ... \, +
x_{\mu-1}}$}. The coordinates $x,y$ run over all sites of a 4-di\-men\-sio\-nal $L^3
\times N_4$ lattice with lattice spacing $a$. The gauge link variables $U_\mu(x)$ are
elements  of the gauge group SU(N).

The U(1)-valued temporal fermionic boundary conditions are  most conveniently
introduced by attaching the boundary phase to the  temporal link on the last
time-slice, $U_4(\vec{x},N_4) \rightarrow  e^{i \varphi} \, U_4(\vec{x},N_4)$
(at the moment $\varphi$ is held fixed and the Fourier transformation of 
(\ref{dualcondensate}) is performed only later in Eq.~(\ref{dressedloops})).
Inserting this into the Dirac operator, we evaluate the propagator in
(\ref{dualcondensate}) for sufficiently  large $m$ as a geometric series
\begin{equation}
\mbox{Tr} \left[ ( m + D_\varphi )^{-1} \right] \; = \; 
\frac{1}{m} \sum_{k=0}^{\infty} \frac{(-1)^k}{m^k} \mbox{Tr} \left[ ( D_\varphi )^k
  \right] \; .
\label{geometric}
\end{equation}
The Dirac operator (\ref{staggeredD}) contains only terms that connect nearest
neighbors. The power $(D_\varphi)^k$ corresponds
to a chain of $k$ hops. The trace in (\ref{geometric}) is over color-
and space-time indices. The latter trace implies that the chains of hops have
to form \emph{closed loops} $l$ (thus on a lattice with even numbers of sites in all 
directions $k$ must be even).
Consequently, the sum in (\ref{geometric}) can be reexpressed as a sum over the set 
${\cal L}$ of all possible closed loops on the lattice,
\begin{equation}
\mbox{Tr} \left[ ( m + D_\varphi )^{-1} \right] =  
\frac{1}{m} \sum_{l\in {\cal L}} \! \frac{ s(l) e^{i \varphi q(l)}}{(2am)^{|l|}}
\, \mbox{Tr}_c \! \! \!
\! \prod_{(x,\mu) \in l} \! \!\! U_\mu(x).  
\label{loopsum}
\end{equation}
The remaining trace Tr$_c$ is over the color indices of the ordered product of all 
link variables $U_\mu(x)$ in a loop $l$, and we use $U_{-\mu}(x) =
U_\mu(x\!-\!\hat{\mu})^\dagger$. By $s(l)$ we denote the sign of a particular loop $l$
which is obtained as product of the staggered sign factors. Each step in the loop
comes with a factor of $1/2am$ from the discretization in  (\ref{staggeredD}) and the
normalization in (\ref{geometric}). The number of steps, i.e., the length of the loop
is denoted by $|l|$. The loops may close around the boundary. When they close around
the temporal boundary, they pick up a factor of $\exp(i\varphi)$ if they run forward in
time, and a factor of $\exp(-i\varphi)$ for backward running. Denoting the number of
times a loop $l$ winds around the compact time direction by its winding number $q(l)
\in \mathds{Z}$, we obtain the factor $\exp(i \varphi q(l))$ in (\ref{loopsum}). 

This is how the boundary angle $\varphi$ can be used to \emph{distinguish between
closed loops of different winding number} \cite{paper1}.  When expression
(\ref{loopsum}) is inserted into the formula (\ref{dualcondensate}) for the dual
condensate, the $\varphi$-integration with the additional Fourier factor  $\exp(-i
\varphi n)$ projects to loops of a particular winding number $n$. We finally obtain
\begin{equation}
\widetilde{\Sigma}_n (m,V)  =  
\frac{1}{Vm} \sum_{l\in {\cal L}(n)}  \frac{ s(l)}{(2am)^{|l|}}
\left\langle \! \mbox{Tr}_c \! \!
\! \prod_{(x,\mu) \in l} \!\! U_\mu(x) \! \right\rangle_G \!\!,  
\label{dressedloops}
\end{equation}
where the sum now runs over the set ${\cal L}(n)$ of loops that wind $n$-times around
the compact time direction. The case of $n=1$, i.e., the dual condensate 
$\widetilde{\Sigma}_1 (m,V)$ which corresponds to loops that wind  exactly once, is
what we refer to as the "dressed Polyakov loop".  From (\ref{dressedloops}) it is
obvious that in the large-$m$ limit the dominant contribution is the conventional thin
Polyakov loop (as this is the shortest loop winding once).

Let us finally discuss the behavior of the dual condensates under a center
transformation, $U_4(\vec{x},t_0) \rightarrow  z \, U_4(\vec{x},t_0)$, where all
temporal links on a time-slice, i.e., at some fixed $t_0$, are multiplied with an
element $z$ of the center of the gauge group. The gauge action and the measure are
invariant under this transformation, but the center symmetry is broken spontaneously at
the critical temperature $T_c$, signaling the transition to the deconfined phase
\cite{znbreaking}. Both the thin Polyakov loop and the dressed Polyakov loop are order 
parameters for center symmetry. More generally a loop
that winds $n$-times around the compact time direction picks up a net factor of $z^n$
and thus our dual condensates transform under a center transformation as
$\widetilde{\Sigma}_n  \rightarrow  z^n  \widetilde{\Sigma}_n$. In particular the
dressed Polyakov loop, i.e., the dual condensate for $n=1$, transforms as
$\widetilde{\Sigma}_1  \rightarrow z  \widetilde{\Sigma}_1$, which is the same
transformation law as for the thin Polyakov loop. 

\vskip4mm
\noindent
{\bf Spectral sums and numerical analysis}
\vskip1mm
\noindent
An elegant way to express the dual condensate (\ref{dualcondensate}) is as a
spectral sum over all Dirac eigenvalues $\lambda^{(i)}_\varphi$ (again evaluated for
boundary angle $\varphi$),
\begin{equation}
\widetilde{\Sigma}_n (m,V) =  
 \int_0^{2\pi} \! \frac{d\varphi}{2\pi}\;
 \frac{e^{-i\varphi n}}{V} \, \sum_i
 \left\langle \big(m + \lambda^{(i)}_\varphi\big)^{-1} 
 \right\rangle_G .
\label{spectralsum}
\end{equation}
First of all this spectral representation consists of a finite sum over the eigenvalues
(due to the regularization on a finite lattice), whereas the representation (\ref{dressedloops})
contains an infinite sum over loops (even on a finite lattice). 

Secondly, the sum on the rhs.\ of (\ref{spectralsum}) can easily be computed
numerically and then allows for the appealing possibility to study the role of
individual parts of the spectrum. Since the (purely imaginary) eigenvalues appear in the
denominator, we expect the dual condensate to be dominated by IR modes (for
small mass $m$) and thus to have a well-defined continuum limit 
(see also the discussion of the convergence of the
spectral sums in \cite{wipf2}). We stress that the above 
discussed properties under center transformations are independent of the mass parameter 
$m$. Naturally one is interested in small values of $m$, where the spectral sums are 
IR dominated. For the Figures 1 and 2 below we use masses near the strange quark mass and 
only in Fig.~3, where we illustrate the transition to the large mass behavior, 
we increase the mass to $m = 1$ GeV.

\begin{figure}[t]
\begin{center}
\includegraphics[height=52mm,clip]{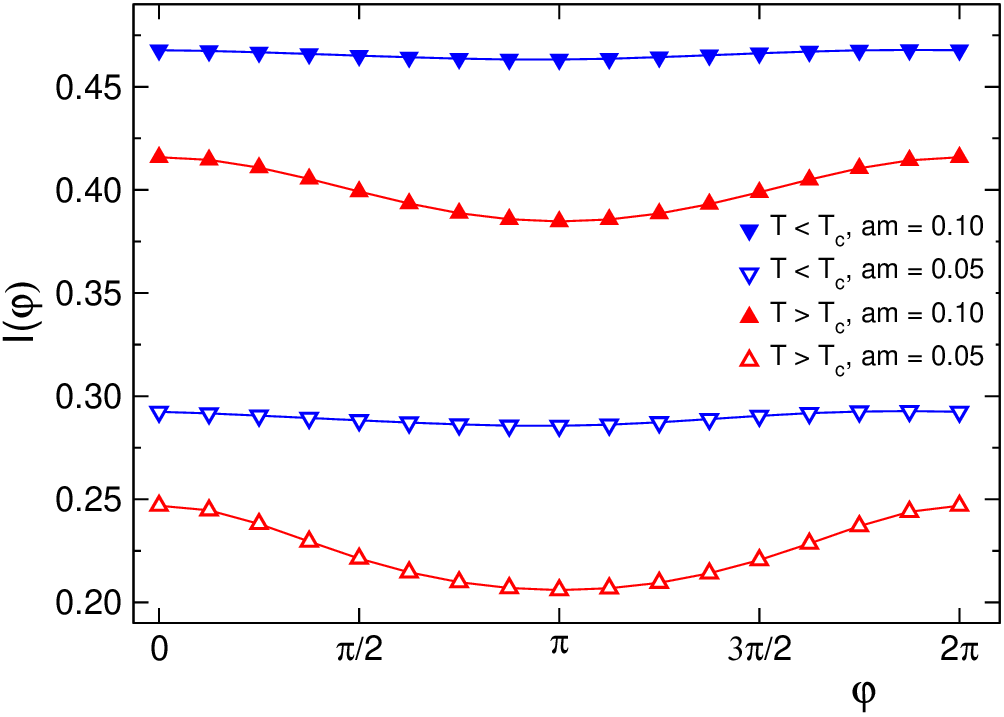} 
\end{center}
\caption{The integrand $I(\varphi) = V^{-1}
\sum_i \big\langle (m + \lambda^{(i)}_\varphi)^{-1}
\big\rangle_G$ of (\ref{spectralsum}) in lattice units
for two values of $am$. The data are from
20 gauge configurations on $12^3 \times 6$ lattices below 
($T$ = 255 MeV, $a$ = 0.129 fm) and above
$T_c$ ($T$ = 337 MeV, $a$ = 0.098 fm).  }
\label{integrandphase}
\end{figure}

For our numerical analysis of the spectral representation (\ref{spectralsum})  we
compute complete spectra of the staggered Dirac operator (\ref{staggeredD}) at
different boundary angles $\varphi$ with parallel LAPACK routines. We use quenched
SU(3) gauge configurations generated from the L\"uscher-Weisz gauge action \cite{LuWe}.
The scale was determined \cite{scale} from the Sommer parameter setting $r_0 = 0.5$
fm.  We use $L^3 \times N_4$ lattices with $L$ ranging from 8 to 14 and $N_4$ from 4
to 8, and adjust the couplings such that we have ensembles below 
and above the critical temperature $T_c \sim$ 300 MeV. For the gauge
configurations above $T_c$ the thin Polyakov loop has a non-vanishing expectation value
and for our numerical study we use configurations where the thin
Polyakov loop is real. Below we discuss how the spectral sum
for the dressed Polyakov loop may generate also the two possible complex phases. 
All errors shown are statistical errors from the Jackknife method.  

\begin{figure}[t]
\begin{center}
\includegraphics[height=52mm,clip]{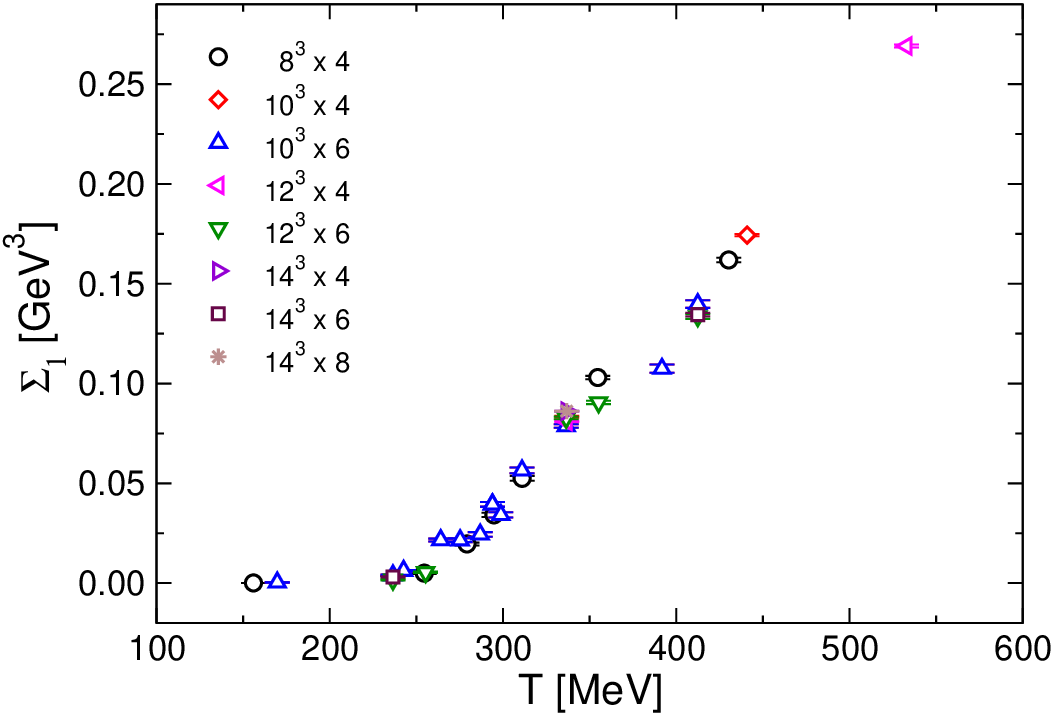} 
\end{center}
\caption{The dressed Polyakov loop at $m = 100$ MeV in units of GeV$^3$ 
as a function of the temperature $T$ in MeV.}
\label{phasetransition}
\end{figure}

We begin with analyzing the $\varphi$-dependence of the integrand in
(\ref{spectralsum}). In Fig.~\ref{integrandphase} we plot this integrand  versus
$\varphi$ for two values of $am$ comparing an ensemble below $T_c$ to one above
$T_c$. It is obvious that below $T_c$ the integrand is essentially constant, while
above $T_c$ it shows a pronounced cosine type of behavior. Integrating over $\varphi$
with the weight $\exp(-i \varphi)$, i.e., $n=1$, gives a vanishing dressed Polyakov
below $T_c$, while above $T_c$ a non-vanishing value is observed.  We conclude that the
transition from confinement to deconfinement leads to a different response of
the spectral sums to the changing temporal fermion boundary conditions. Below we will
demonstrate that the IR modes play the dominant role in this process. 

Above $T_c$, when using instead of an ensemble with real thin Polyakov loop,   
configurations where the thin Polyakov loop has one of the two complex phases, 
the integrand is shifted by $\pm 2\pi/3$, but otherwise has the same
form as depicted in Fig.~\ref{integrandphase}. Integrating this shifted integrand 
shows that the dressed Polyakov loop produces the same $\mathds{Z}_3$  phase pattern as
the thin Polyakov loop. 

Now we demonstrate that the dressed Polyakov loop does indeed signal the phase
transition. In Fig.~\ref{phasetransition} we show the results for the dressed Polyakov
loop at $m = 100$ MeV as a function of $T$. The necessary $\varphi$-integration  was
implemented with the extended Simpson rule using typically 8 or 16 values of $\varphi$. It is
obvious that below $T_c$ the dressed Polyakov loop vanishes, while above $T_c$ it assumes
a non-vanishing value, signaling that the center symmetry is broken in the deconfining
phase.

Another interesting connection between quark condensate and dressed Polyakov
loop is obtained by using the Banks-Casher type of representation already addressed
above. After performing the consecutive limits of infinite volume and vanishing mass,
the chiral condensate can be written as the density of eigenvalues at the origin also
for arbitrary boundary angle $\varphi$. The dual condensate is then obtained by
integrating the $\varphi$-dependent spectral density $\rho(0)_\varphi$,  and for the
case of $n=1$ we find
\begin{equation}
\widetilde{\Sigma}_1 \; = \; \int_0^{2\pi} 
\frac{d \varphi}{2} \, e^{-i \varphi} 
\rho(0)_\varphi \; .
\label{loopspecdens}
\end{equation}
Below $T_c$ the spectral density at the origin is constant as a function of $\varphi$
and a vanishing dressed Polyakov loop emerges. More interesting is the situation above
$T_c$, where a non-trivial $\varphi$ dependence is necessary for a non-vanishing dressed
Polyakov loop. Naively one would think that above $T_c$ the spectral density at the
origin must be zero, such that the chiral condensate may vanish. However, in
\cite{gapphase} (for a different phase convention)
it was shown that the spectral gap depends on the relative phase between
the boundary angle $\varphi$ and the phase $\theta$ of the Polyakov loop. If 
$\varphi$ equals the negative Polyakov loop phase the gap closes completely,
giving rise to a non-zero spectral density. Inserting $\rho(0)_\varphi \propto
\delta(\varphi + \theta)$ in 
(\ref{loopspecdens}) one obtains a non-vanishing dressed Polyakov loop above $T_c$ with
the correct phase $\theta$.

\begin{figure}[t]
\begin{center}
\includegraphics[width=87mm,clip]{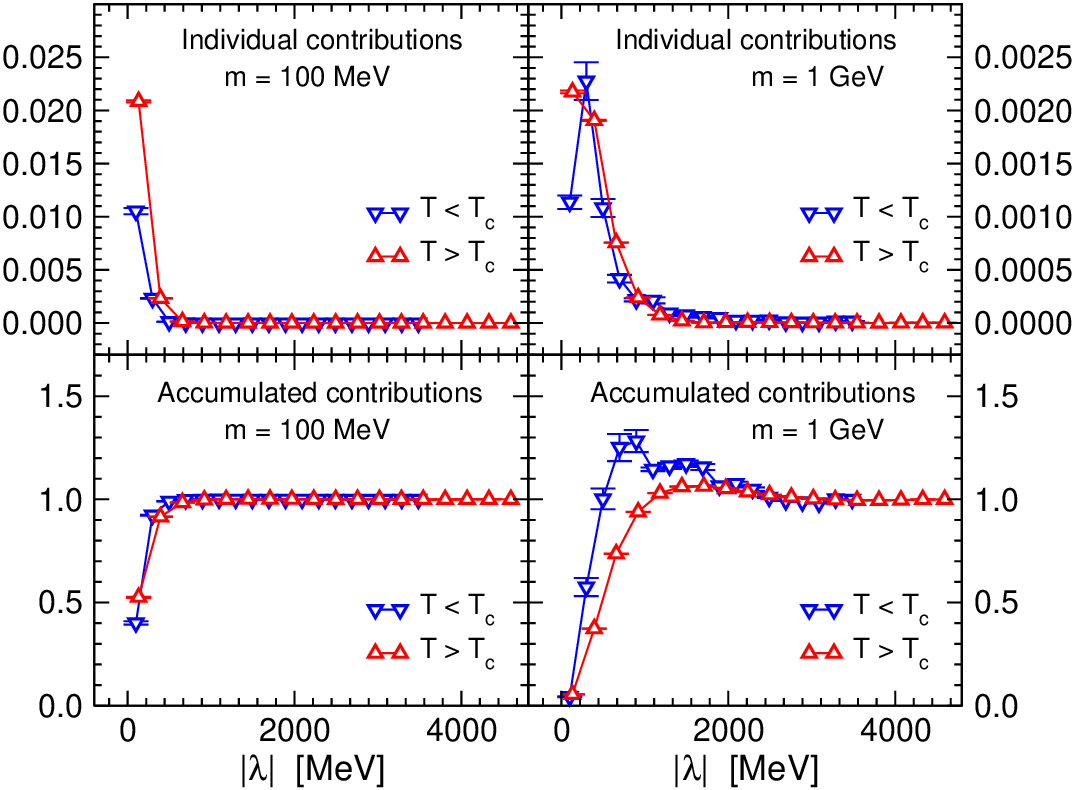} 
\end{center}
\caption{The upper plots show the (normalized) individual contributions 
$|C(|\lambda|)/\widetilde{\Sigma}_1|$  
to (\ref{spectralsum}) versus $|\lambda|$ for two values of $m$, while the lower plots
are for the normalized accumulated contributions $|A(|\lambda|)/\widetilde{\Sigma}_1|$ as
a function of $|\lambda|$. The data are from the same ensembles already used in Fig.~1.}
\label{contribaccum}
\end{figure}

Let us now consider the individual contributions $C(\lambda) = (2\pi V)^{-1} \int d \varphi
\exp(-i\varphi) \big\langle (m + \lambda_\varphi)^{-1} \big\rangle_G$ to the spectral
sum (\ref{spectralsum}). The upper plots of Fig.~\ref{contribaccum} show  
$|C(\lambda)/\widetilde{\Sigma}_1|$ as a function of $|\lambda|$. Since the case 
of vanishing quark mass $m$
can be obtained only through a limiting procedure, we compare two different values of
$m$. In both cases the largest contributions come from the IR end of the spectrum.
However, for the smaller mass the contributions beyond the deep IR have died out
completely.

When considering the relative role of IR and UV contributions, one must take into account that
the density of eigenvalues increases strongly with increasing  $|\lambda|$. Thus in the lower
plots of Fig.~\ref{contribaccum} we show the accumulated  contribution $A(|\lambda|) =
\sum_{|\lambda^\prime| \leq |\lambda|} C(\lambda^\prime)$. In particular we  plot the
normalized function $|A(|\lambda|)/\widetilde{\Sigma}_1|$ which approaches 1 in the UV
limit. Here the difference between the large and the small mass is rather pronounced: For the
large mass, where the dressed loop is closer to the thin Polyakov loop, the accumulated IR
modes overshoot the value of 1 and the UV contributions are necessary to get the correct value.
This is similar to the observations in  \cite{paper2,paper3,soeldner} for the thin Polyakov
loops. Towards the chiral limit, however, the approach to the final value of 1 becomes
monotonic, and only the IR modes give sizable contributions to the spectral sum for the dressed
Polyakov loop. The scale up to which eigenvalues are relevant grows with $T$.
\\
\\
\noindent
{\bf Concluding remarks}
\vskip1mm
\noindent
In this letter we have shown that a duality transformation of the quark condensate with
respect to the fermionic temporal boundary condition gives rise to an order parameter
for center symmetry. This order parameter can be viewed as a set of closed loops with
the same winding number around compact time. For the case of single winding, which we
refer to as the dressed Polyakov loop,  our observable interpolates between the quark
condensate (via a Fourier transform) and the thin Polyakov loop, in the limits of
vanishing and infinite mass, respectively.

We studied the corresponding spectral sums of Dirac operator eigenvalues numerically
for quenched gauge configurations. It was shown that the transition from the confined
to the deconfined phase is seen as a different dependence of the spectral sums on the
fermionic boundary condition. Decomposing the spectral sum shows that the main signal  
comes from the IR part of the spectrum.

As an outlook we stress that our theoretical considerations also hold for other gauge
groups (where, e.g., for SU(2) the phase transition can be of different order) and for
the case of dynamical quarks, the consequences of which would be interesting to study
numerically.

\begin{acknowledgments}
We thank C.B.\ Lang, K.\ Langfeld, W.\ S\"oldner, T.\ Kovacs, P.\ van Baal,
J.\ Verbaarschot and A.\ Wipf for valuable discussions. The numerical analysis was
done at the ZID, University of Graz. E.B.\ and C.G.\
are supported by NAWI-GASS and the FWF (DK W1203-N08 and P 20330-N16) and 
F.B.\ by DFG (BR 2872/4-1).   
\end{acknowledgments}

\end{document}